\documentclass[aps,prb,preprint,groupedaddress,showpacs,amsmath,amssymb]{revtex4}

\usepackage{graphicx}% Include figure files
\usepackage{dcolumn}% Align table columns on decimal point
\usepackage{bm}% bold math
\usepackage{hyperref}

\begin{document}

\title{Tuning of charge density wave transitions in LaAu$_x$Sb$_2$ by pressure and Au - stoichiometry.}

\author{ Li Xiang,$^{1,2}$  Dominic H. Ryan,$^{1,2,3}$ Warren E. Straszheim,$^1$  Paul C. Canfield,$^{1,2}$ and Sergey L. Bud'ko$^{1,2}$}

\affiliation{$^{1}$Ames Laboratory, US DOE, Iowa State University, Ames, Iowa 50011, USA}
\affiliation{$^{2}$Department of Physics and Astronomy, Iowa State University, Ames, Iowa 50011, USA}
\affiliation{$^{3}$Physics Department and Centre for the Physics of Materials, McGill University, 3600 University Street, Montreal, Quebec, H3A 2T8, Canada}

\date{\today}

\begin{abstract}
Two charge density wave transition can be detected in LaAuSb$_2$ at $\sim 110$ and $\sim 90$ K by careful electrical transport measurements.  Whereas control  of the Au site occupancy in LaAu$_x$Sb$_2$ (for $0.9 \lesssim x \lesssim 1.0$) can suppress each of these transitions by $\sim 80$~K, the application of hydrostatic pressure can completely suppress the lower transition by $\sim 10$~kbar and the upper transition by $\sim 17$~kbar.  Clear anomalies in the resistance as well as the magnetoresistance are observed to coincide with the pressures at which the charge density wave transitions are driven to zero. 
\end{abstract}

%\pacs{}
%\keywords{}

\maketitle

\section{Introduction}

Charge density wave (CDW) phenomena continue to attract the attention of condensed matter physicists. \cite{joh08a,eit13a,zhu15a,pou16a,sai19a} One of the fascinating research lines continues to be coexistence and competition of CDW and other collective phenomena like superconductivity. \cite{bil76a,bal79a,mac81a,gab01a,wan18a}  On a more basic level though, despite more than half a century history, \cite{pei30a,pei55a,fro54a} the driving forces for CDW formation in different materials as well as classification of CDWs are still under discussion. \cite{joh08a,eit13a,zhu15a} In this context, the identification and studies of new CDW materials are important to diversify the pool of well studied systems.

Ternary LaAgSb$_2$ \cite{bry95a,sol95a} is a non-magnetic member of the family of tetragonal (ZrCuSi$_2$ structure type, space group $P4/nmm$) RAgSb$_2$ (R = rare earth) compounds  with diverse physical properties. \cite{sol95a,mye99a} The anomalies in many physical properties \cite{mye99a,son03a,lue07a,bud08a,mun11a,che17a} suggested formation of two CDWs in LaAgSb$_2$, one at $T_{CDW1} = 208$~K and another at  $T_{CDW2} = 186$~K. X-ray diffraction studies have provided direct evidence of CDW formations below $T_{CDW1}$ (along the $a$ direction with $\tau_1 = 0.026(2\pi/a)$) and below $T_{CDW2}$ (along the $c$ direction with $\tau_2 = 0.16(2\pi/c)$). \cite{son03a}. Moreover, LaAgSb$_2$ was suggested to have Dirac states with a close relationship between the Dirac cone to the CDW ordering. \cite{wan12a,shi16a}

Effects of rare earth substitution \cite{bud06a,mas14a} and hydrostatic pressure \cite{bud06a,tor07a} on the temperature of the higher CDW transition, $T_{CDW1}$ were studied. The observed suppression of the $T_{CDW1}$ was explained as a combination of increase of the 3-dimensional character of LaAgSb$_2$ (decrease of $c/a$ value) and the substitution - related disorder. It was also recognized that  the pressure response of  $T_{CDW1}$ could be affected by local moment magnetism as well as hybridization due to rare earth  (R = Ce, Pr, Nd, ...) substitution  in (La$_{1-x}$R$_x$)AgSb$_2$. \cite{tor07a}

Less than a decade ago, CDW formation at $\sim 95$~K  was reported, even if in passing, for a closely related material, LaAuSb$_2$. \cite{seo12a} This work was followed by recent (magneto-)  electrical and thermal transport studies, as well as ultrafast pump-probe spectroscopy measurements, on LaAuSb$_2$ \cite{kuo19a} providing  evidence for partial gapping of the Fermi surfaces during the CDW transition at around 88~K. Additionally, resistivity data for several samples of the La(Ag$_{1-x}$Au$_x$)Sb$_2$ series were reported, \cite{mas14a} indicating suppression of the CDW transition with Au substitution (only one CDW transition was detected). In this work the Au-end compound had $T_{CDW} = 88$~K and was designated as LaAu$_{0.88}$Sb$_2$.

Indeed, in contrast to the stoichiometric RAgSb$_2$ series, \cite{mye99a,mye99b} the RAuSb$_2$ family (in particular more thoroughly studied CeAuSb$_2$) was suspected to have transition metal deficiency \cite{seo12a,zha16a}. Different Au occupancies, $x$, on the single Au site, Au $2b$, are most probably responsible for different values of the observed $T_{CDW}$ as well as the range of residual resistivity ratios, RRR, in LaAu$_x$Sb$_2$. \cite{seo12a,kuo19a,mas14a}

Having in mind apparent similarity between LaAu$_x$Sb$_2$ and the much more studied LaAgSb$_2$ as well as the additional complexity of the former material due to its transition metal deficiency, in this work we aim to address several questions: (a) can we tune and control Au occupancy $x$ in  LaAu$_x$Sb$_2$? (b) as in the case of LaAgSb$_2$, is there a second, lower temperature CDW in  LaAu$_x$Sb$_2$? (c)  how are the CDW transitions affected by $x$? (d) is the pressure response of the CDW transitions in  LaAu$_x$Sb$_2$ different from those in  LaAgSb$_2$? (e) will we be able to suppress the CDW transitions in  LaAu$_x$Sb$_2$ completely, and if so, are there any anomalies in (magneto-) transport associated with the CDW quantum phase transitions / quantum critical points?

\section{Experimental details}

Single crystals of LaAu$_x$Sb$_2$  were grown from an antimony-rich self-flux following the method described in Refs. [\onlinecite{mye99a,zha16a}]. 5--6 grams of the pure elements  were loaded into the lower (growth) half of an alumina Canfield crucible set \cite{can16a} which was capped with an alumina frit and a second (catch) alumina crucible. This assembly was loaded into an amorphous silica tube, evacuated, back-filled to $\sim$150~mbar with argon, and sealed. The sealed tubes were heated to 1050~$^{\circ}$C over a period of 10~hours, held for 8~hours to ensure the formation of a homogeneous liquid, then cooled to 800~$^{\circ}$C over a period of 10~hours prior to starting the crystal growth (we found no solids down to 750~$^{\circ}$C). Crystal growth occurred during the 100~hour cooling from 800~$^{\circ}$C to 670~$^{\circ}$C, after which the excess flux was removed using a centrifuge. Typical yields were 500~mg--600~mg as 1--3 well-faceted crystals, $\sim$5~mm on each side and $\sim$1~mm thick.

To investigate whether reported Au deficiency \cite{seo12a,mas14a} was affected and could be tuned by the initial stoichiometry of the melt, five initial La :  Au : Sb growth compositions were used: 1 : 1 : 20 (denoted Au1), 1 : 2 : 20 (Au2), 1 : 4 : 20 (Au4),  1 : 6 : 20 (Au6), and 1 : 8 : 20 (Au8). Similar excess of Au was found to yield near stoichiometric (as measured by energy dispersive spectroscopy) CeAuSb$_2$; initial Ce : Au : Sb ratios of 1 : 6 : 12~\cite{zha16a} gave residual resistivity ratios between 6 and 9, quite similar to our findings for Au6 and Au8 growths as shown below. The excess Au in the starting melt did not appear to adversely affect the size of the final crystals, however we observed slightly more surface contamination by flux (a mixture of Sb and AuSb$_2$) in the higher Au derived samples. 

Cu-K$_{\alpha}$ x-ray diffraction patterns were taken for all of the samples using a Rigaku Miniflex-II diffractometer. For each starting growth composition the crystal was cleaned to remove any residual flux (mechanical scraping followed by wiping with an ethanol-soaked paper tissue) then a small piece was broken off and hand ground under ethanol to minimize oxidation. The powder was mounted on a low-background single-crystal silicon plate using a trace amount of Dow Corning silicone vacuum grease. The mount was spun during data collection to reduce possible effects of texture. Data taken for Rietveld refinement were collected in two overlapping blocks $10^{\circ} \leq 2\theta \leq 48^{\circ}$ and $38^{\circ} \leq 2\theta \leq 100^{\circ}$, with the second block counted for 4-5 times longer than the first to compensate for the loss of scattered intensity at higher angles due to the x-ray form factors. This strategy typically yielded $\sim$10 Bragg reflections with intensities over 2000~counts and many other statistically significant reflections out to $2\theta$ = 100$^{\circ}$ allowing us to decouple the effects of site occupation and thermal factors in the structural analysis. The diffractometer and analysis procedures were checked using $\rm Al_2O_3$ (SRM 676a\cite{nist}) and our fitted values of $a=$4.7586(2)~\AA\ and $c=$12.9903(7)~\AA\ were both 1.6(4)$\times10^{-4}$ smaller than the values on the certificate\cite{nist}, suggesting a small but statistically significant mis-calibration of the instrument. The fitted lattice parameters given in the analysis that follows do not include this correction.

Standard, linear 4-probe ac resistivity was measured on bar - shaped samples of  LaAu$_x$Sb$_2$ in two arrangements: $I || ab$ and, when needed, $I || c$. The frequency used was 17 Hz, typical current values were 3 mA for in-plane electrical transport and 5 mA for the $c$ - axis measurements. Magnetoresistance was measured in a transverse configuration: $H || c$  for the in-plane transport and $H || ab$ for $I || c$. The measurements were performed using the ACT option of a  Quantum Design Physical Property Measurement System (PPMS).

For selected samples, resistivity measurements under pressure were performed in a hybrid, BeCu / NiCrAl piston - cylinder pressure cell (modified version of the one used in Ref. \onlinecite{bud86a}) in the temperature and magnetic field environment provided by a PPMS instrument. A 40 : 60 mixture of light mineral oil and n-pentane was used as a pressure-transmitting medium.This medium solidifies at room temperature in the pressure range of 30 - 40 kbar, \cite{bud86a,kim11a,tor15a} which is above the maximum pressure in this work. Elemental Pb was used as a low temperature pressure gauge.\cite{eil81a} It has been shown \cite{its64a,its67a,tho84a} that in piston-cylinder pressure cells high temperature pressures are different from low temperature pressures. Given that the transitions of interest occur below $\sim 115$~K (see below), here we simply use the Pb gauge pressure value. This may give rise to pressure differences with the values determined by Pb gauge by at most few tenths of a kbar.

Chemical analysis of the crystals was performed using an Oxford Instruments energy-dispersive x-ray spectroscopy (EDS) system on a Thermo Scientific Teneo scanning electron microscop.  The measurements were performed on polished $ab$ surfaces of single crystals with 4 points taken for every sample.

\section{Results}

\subsection{Tuning and control of Au concentration in LaAu$_x$Sb$_2$}

\begin{figure}
\begin{center}
\includegraphics[angle=0,width=200mm]{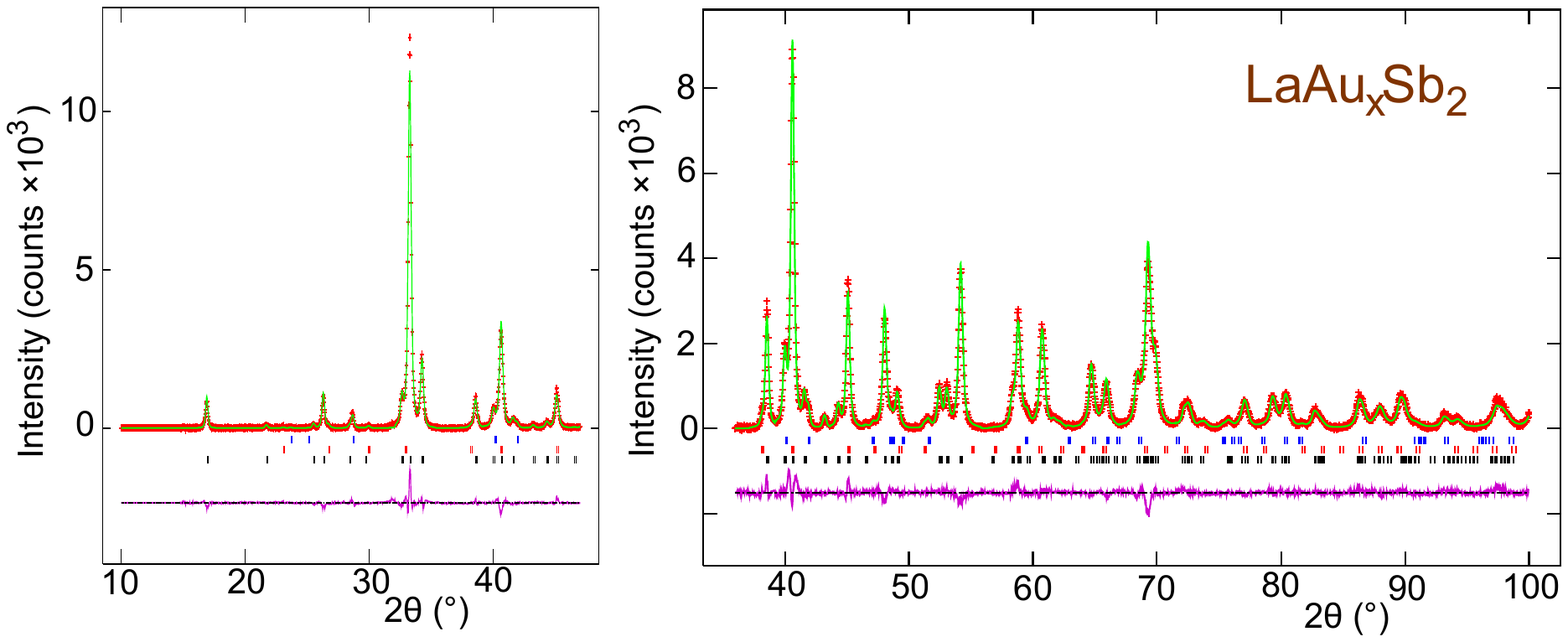}
\end{center}
\caption{(color online) Cu-K$_{\alpha}$ x-ray diffraction patterns for the Au4 growth of LaAu$_x$Sb$_2$ showing the two overlapping data blocks that were co-fitted using the GSAS/EXPGUI packages. \cite{lar00a,tob01a} The red points are the data and the green lines show the fits with the residuals shown below each fitted pattern. The Bragg markers show the positions of the reflections from (top) Sb, (middle) AuSb$_2$ and (bottom) LaAu$_x$Sb$_2$. } \label{F1}
\end{figure}

The x-ray diffraction patterns were fitted using the GSAS/EXPGUI packages \cite{lar00a,tob01a}. Small amounts of residual flux were generally observed as impurity phases and were included in the fits as necessary. Fig.~\ref{F1} shows a typical x-ray diffraction data set for the Au4 growth of LaAuSb$_2$ with 1.7~wt.\% AuSb$_2$ and 3.2~wt.\% Sb as impurities. In the fit, the occupation of the Au 2$b$ site was allowed to vary and was found to be less than one (see below), whereas the occupations of La, Sb1 and Sb2 sites were fixed as 1. The results of the Rietveld analysis of the powder x-ray data for all 5 LaAu$_x$Sb$_2$ samples are listed in Tables \ref{T1} and \ref{T2} in the Appendix A. 

The EDS results for the 5 LaAu$_x$Sb$_2$ samples are presented in Table \ref{T3} in the Appendix A. The values in the table are the average of the measurements taken at 4 different places on the samples' surfaces, standard deviations are listed in the parentheses. 

\begin{figure}
\begin{center}
\includegraphics[angle=0,width=140mm]{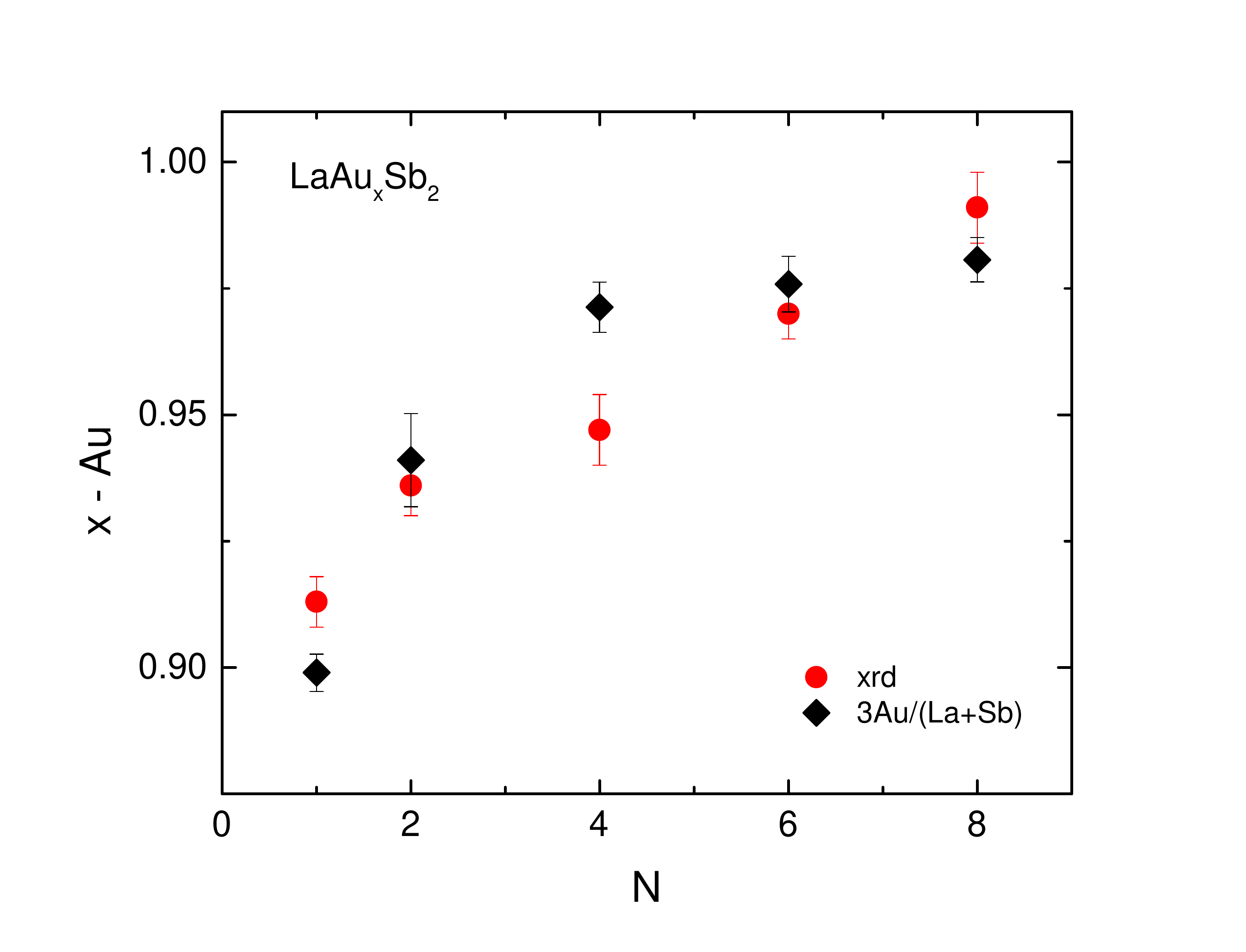}
\end{center}
\caption{(color online)  The fitted occupation of the Au site (red circles) in LaAu$_x$Sb$_2$ as a function of N, in starting stoichiometry 1 (La) : N (Au) : 20 (Sb). plotted together with Au  concentration relatively to (La + Sb) (black diamonds) determined from EDS measurements. } \label{F2}
\end{figure}

\begin{figure}
\begin{center}
\includegraphics[angle=0,width=140mm]{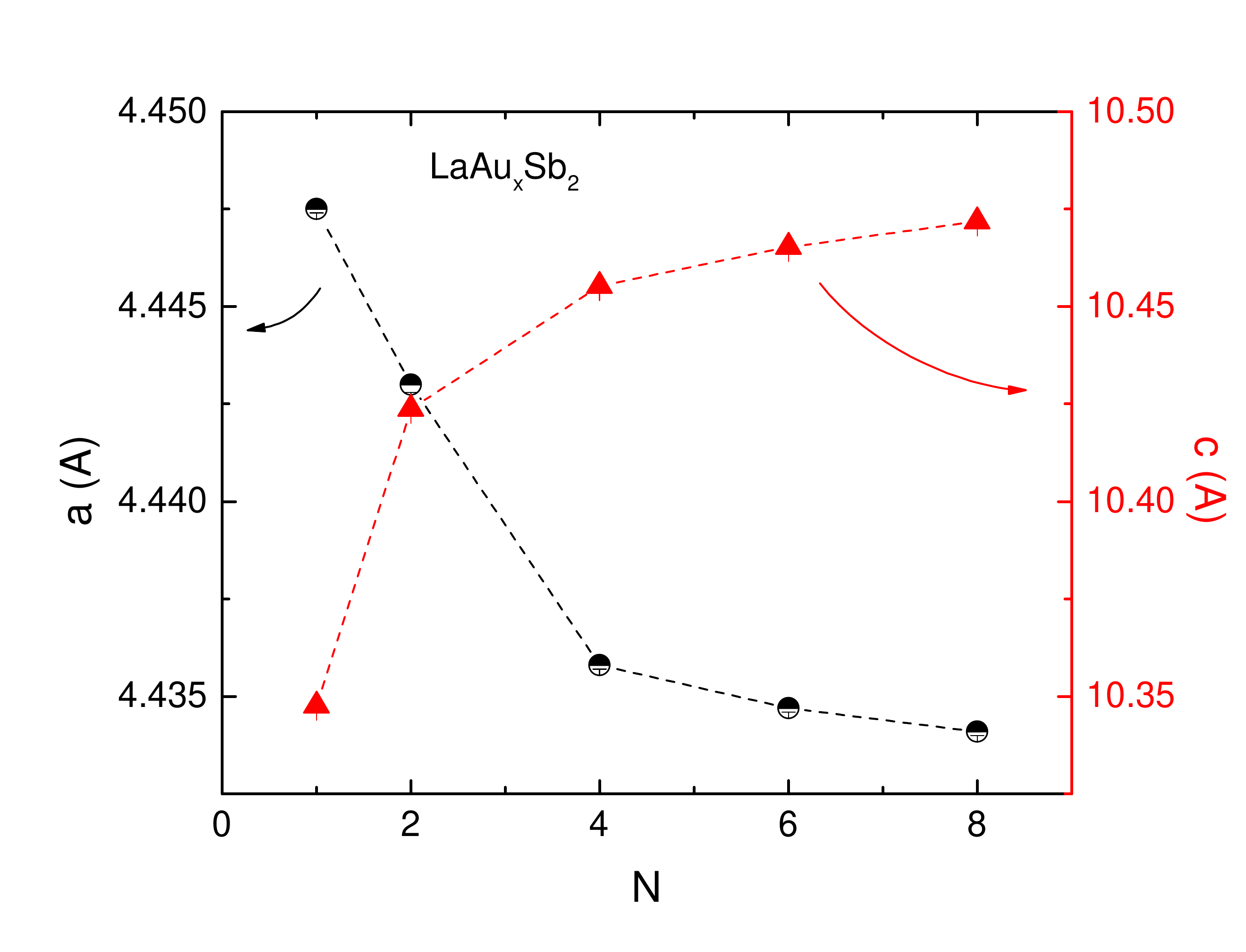}
\end{center}
\caption{(color online) Lattice parameters $a$ and $c$ of  LaAu$_x$Sb$_2$ plotted as a function of N, in starting stoichiometry 1 (La) : N (Au) : 20 (Sb). } \label{F3}
\end{figure}

Analysis of the x-ray diffraction as well as EDS results clearly show that increasing the gold content of the starting mixture had a significant effect on the Au occupancy of the grown crystals. The fitted occupation of the Au 2$b$ site increased from 0.913(5) for Au1 to 0.991(7) by Au8. This span of Au concentrations is consistent with the EDS data. The Au occupancy from the x-ray diffraction and atomic ratio of 3Au/(La+Sb) from the EDS data are plotted together in Fig. \ref{F2}. The most increase of the Au - concentration appear to happen between the  Au1 and Au4 samples, followed by almost saturation for Au6 and Au8. This saturation effect is also clearly visible in the behavior of the lattice parameters shown in Fig.~\ref{F3}. While we do observe vacancies on the Au site, the Au occupancy determined in this study  is above the 0.88 reported by Masubuchi  {\it et al.} \cite{mas14a} for the Au2 sample that should correspond to the composition that they used. Our Au6 and Au8 samples are within composition range reported for CeAu$_x$Sb$_2$ grown with Ce : Au of 1 : 6. \cite{zha16a}.

In the rest of the text we will use the notation LaAu$_x$Sb$_2$ with $x$ determined from the Rietveld refinement of the powder x-ray data. 

\subsection{Ambient pressure electrical transport and CDW transitions}

\begin{figure}
\begin{center}
\includegraphics[angle=0,width=140mm]{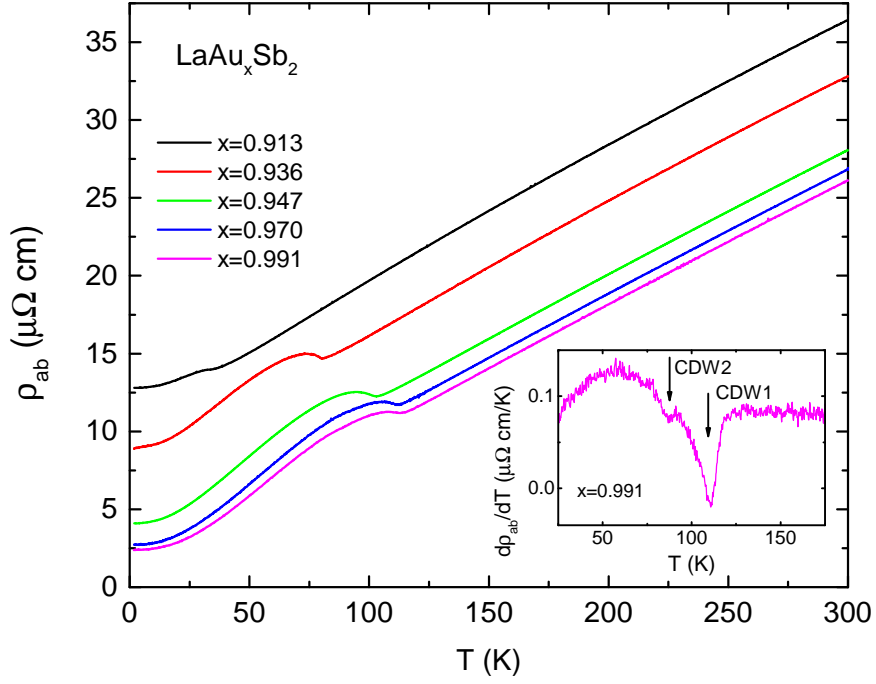}
\end{center}
\caption{(color online) Temperature-dependent, in-plane, resistivity data for the $x = 0.970$ sample together with  resistance data  for the other LaAu$_x$Sb$_2$ samples normalized  to that of the $x =0.970$  sample so that the room temperature slope of the $\rho_{ab}(T)$ data match (see text for details).  Inset: temperature derivative of resistivity for LaAu$_{0.991}$Sb$_2$, arrows mark two CDW transitions. } \label{F4}
\end{figure}

\begin{figure}
\begin{center}
\includegraphics[angle=0,width=140mm]{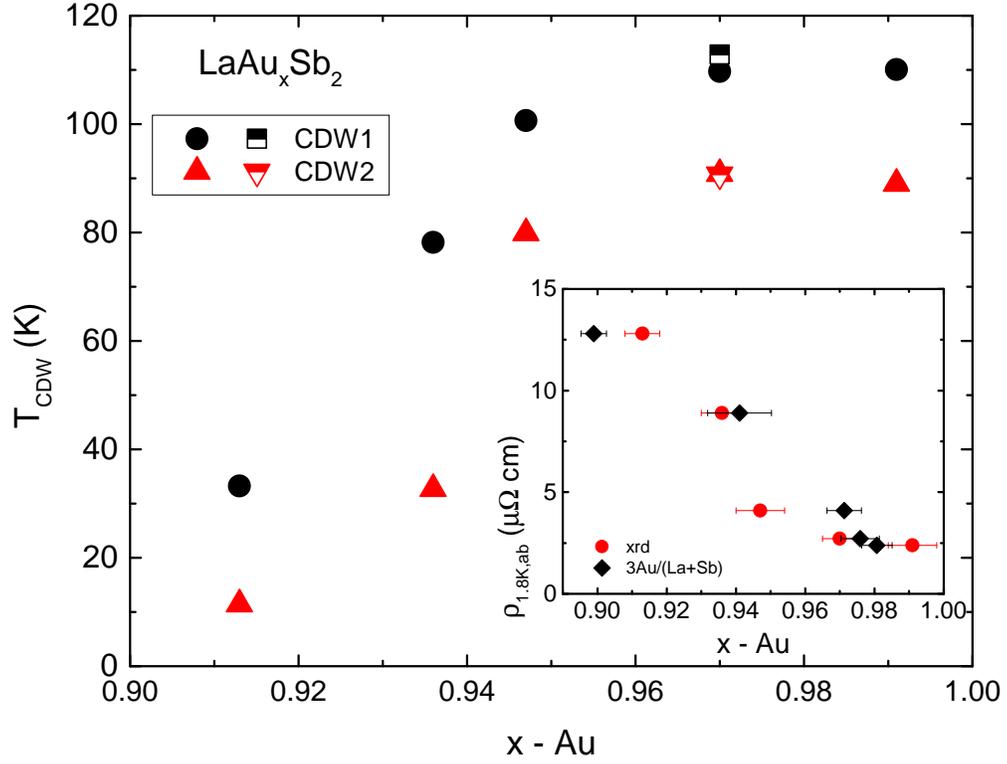}
\end{center}
\caption{(color online) CDW transition temperatures, $T_{CDW1}$ and $T_{CDW2}$ as a function of $x$ determined by Rietveld refinement in  LaAu$_x$Sb$_2$ . Filled and half-filled symbols - from in-plane and $c$-axis resistivity data respectively. Inset: residual resistivity, $\rho_{1.8 \textrm{K},ab}$ as a function of $x$ determined from x-ray diffraction (red circles) and EDS (black diamonds). } \label{F5}
\end{figure}

Temperature-dependent, in-plane, resistivity data for the $x = 0.970$ (N=6) sample is shown in Fig. \ref{F4}.  Given the uncertainty in geometric factors of the resistance bars, we have normalized the resistance data (multiplicatively) for the other LaAu$_x$Sb$_2$ samples to that of the $x =0.970$ (N=6) sample so that the room temperature slope of the $\rho_{ab}(T)$ data match.  This normalization is premised on the anzatz that small changes in Au occupancy will not change the phonon spectra at 300~K (i.e. the electron phonon scattering that dominates the temperature dependent resistivity at 300~K) in any significant manner.  The data shown in Fig. \ref{F4} preserve the RRR values and also demonstrate very clear Matthiessen's rule offsets of the higher temperature ($T > T_{CDW}$) resistivity.  Remarkably, the $\rho_0$ values (inset to Fig. \ref{F5}) vary  as $\sim 1$~$\mu \Omega$~cm per percent Au vacancy; this is consistent with the very gross, textbook rule of thumb \cite{kit05a} for residual resistivity given for generic metallic samples.  We clearly detect two CDW transitions, a higher temperature CDW1 and a lower temperature CDW2.  CDW1 is easily identified in in-plane resistivity measurements (Fig. \ref{F4}). For some of the Au concentrations the lower, CDW2 transition, is seen in the temperature derivatives of the $\rho_{ab}$, as shown, for example in the inset to Fig. \ref{F4}.  For e.g. $x = 0.970$ $I || c$, $\rho_c(T)$, measurements were performed to observe CDW2 more clearly. As was previously observed for pure LaAgSb$_2$, \cite{mye99a,son03a}  by the combination of in-plane and $c$ - axis resistivity measurements, two CDW transitions were detected for all five $x$ - concentrations in LaAu$_x$Sb$_2$.

The CDW temperatures of of LaAu$_x$Sb$_2$ , $T_{CDW1}$ and $T_{CDW2}$, are plotted in Fig. \ref{F5} as a function of the Au site occupancy $x$ determined from the Rietveld refinement of the powder x-ray diffraction spectra. As it was observed for the lattice parameters, the main change in the CDW temperatures happens between $x = 0.913$ and $x = 0.947$ (samples Au1 - Au4).

\subsection{Electrical transport under pressure, CDW quantum critical point}

For measurements of electrical properties under pressure we have chosen  LaAu$_{0.970}$Sb$_2$ as a sample with Au  site almost fully occupied. (For comparison, in Appendix B, we present similar data from measurements on  LaAu$_{0.936}$Sb$_2$. This composition is similar in growth conditions and the value of $T_{CDW1}$ to the samples reported in the literature at ambient pressure. \cite{mas14a,kuo19a})

\begin{figure}
\begin{center}
\includegraphics[angle=0,width=120mm]{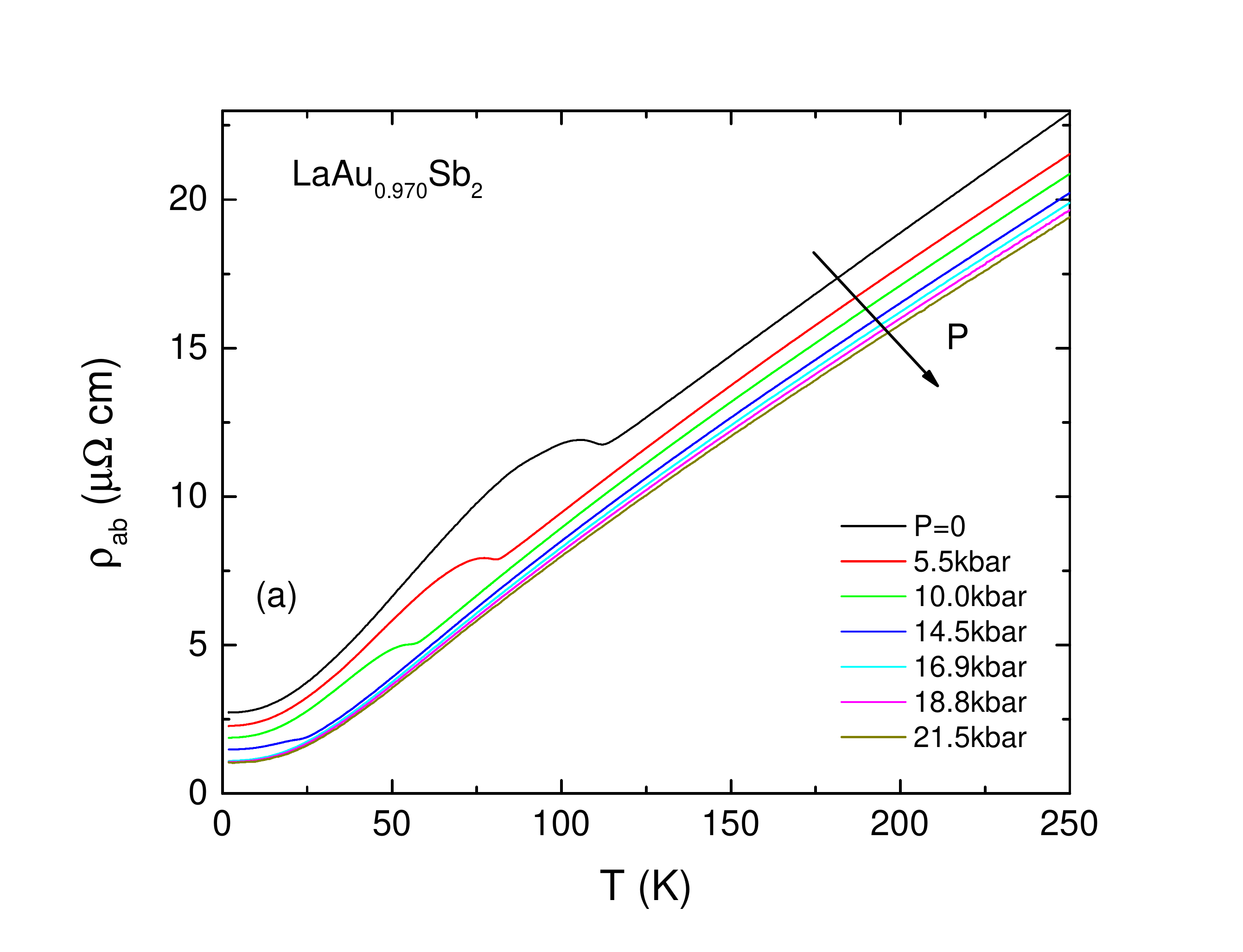}
\includegraphics[angle=0,width=120mm]{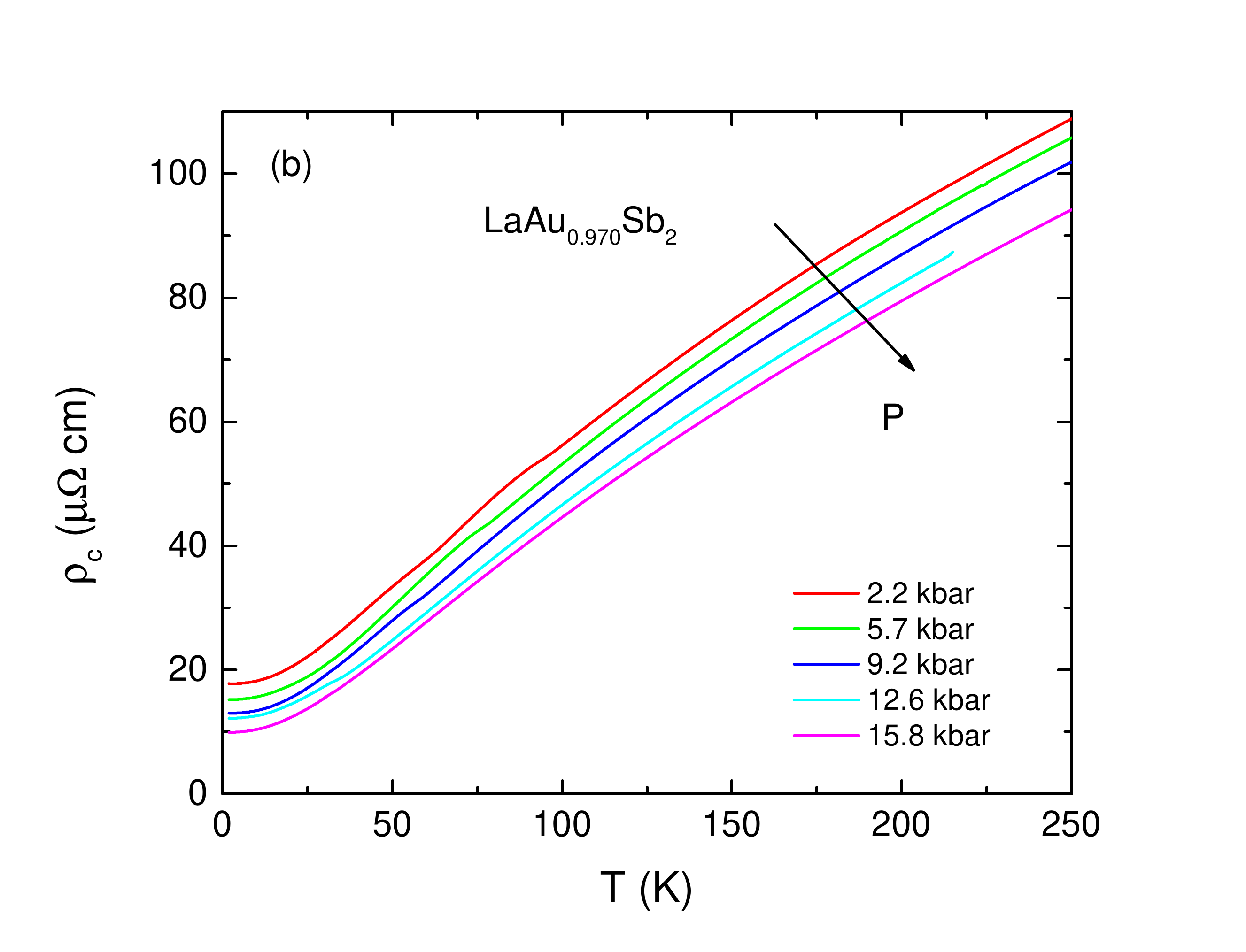}
\end{center}
\caption{(color online) Temperature-dependent (a) in-plane, (b) $c$ - axis resistivity of  LaAu$_{0.970}$Sb$_2$ measured at different pressures. Arrows point to the direction of pressure increase.} \label{F6}
\end{figure}

\begin{figure}
\begin{center}
\includegraphics[angle=0,width=140mm]{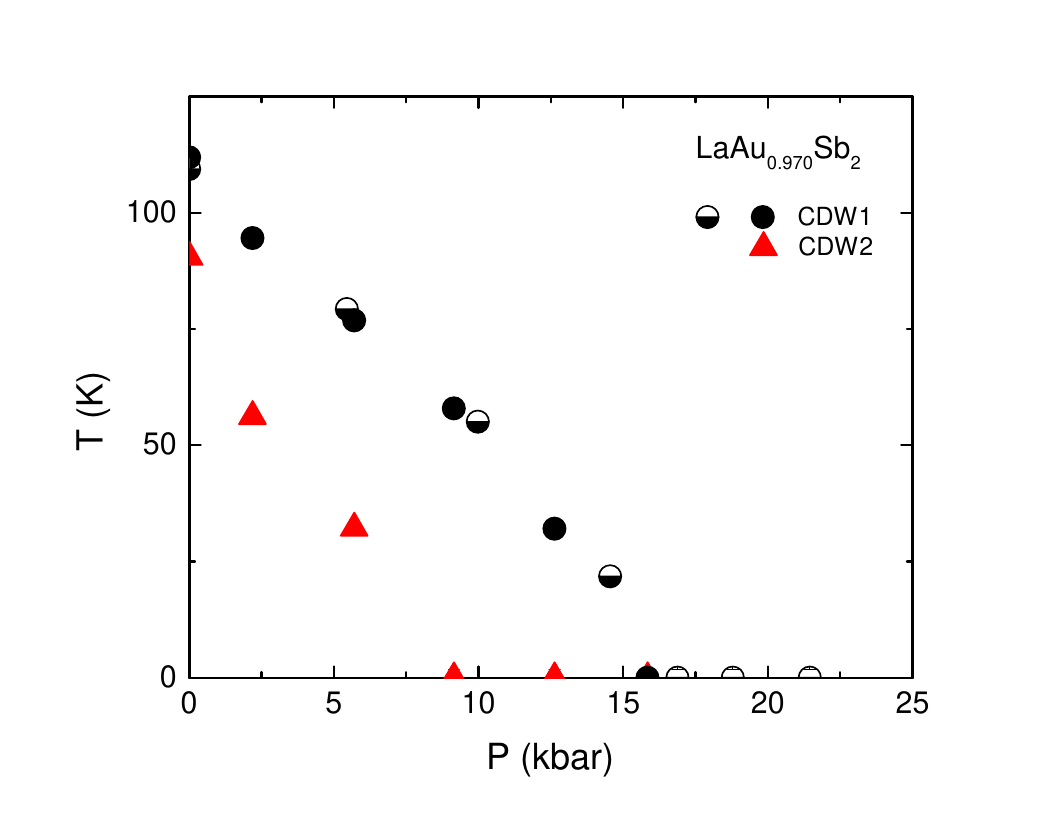}
\end{center}
\caption{(color online) Pressure - temperature phase diagram of  LaAu$_{0.970}$Sb$_2$. Half-filled and filled symbols are from $I || ab$ and $I || c$ runs respectively. Symbols at $T = 0$ correspond to pressures at which no anomalies were detected above 1.8~K. } \label{F7}
\end{figure}

Fig. \ref{F6}(a) shows in-plane resistivity of  LaAu$_{0.970}$Sb$_2$ measured at different pressures up to 21.5~kbar. As in the case of LaAgSb$_2$, \cite{bud06a,tor07a} resistivity decreases under pressure and the feature associated with the CDW1 becomes smaller and shifts down in temperature. For the  LaAu$_{0.970}$Sb$_2$  sample the available pressure range is enough to suppress the $T_{CDW1}$ completely to zero. It should be noted that in this case the  in-plane resistivity measurements (Fig. \ref{F6}(a)) did not present a clear feature for CDW2, so a second series of pressure runs, using $c$-axis resistivity measurements, were performed (Fig. \ref{F6}(b)). In these $\rho_c$ data, both transitions were detected. The $P~-~T$ phase diagram is shown in Fig. \ref{F7}. The suppression of both CDWs is close to linear in pressure. The evaluated  pressure derivatives are $dT_{CDW1}/dP = - 6.0 (2)$~K/kbar and $dT_{CDW2}/dP = - 10 (2)$~K/kbar, and extrapolated critical pressures are $\sim 17$~kbar and $\sim 10$~kbar for CDW1 and CDW2 respectively. For  LaAu$_{0.970}$Sb$_2$ the higher temperature CDW is suppressed somewhat faster than for LaAgSb$_2$, where the pressure derivative value of $-4.3(1)$~K/kbar has been reported. \cite{bud06a,tor07a}.

\begin{figure}
\begin{center}
\includegraphics[angle=0,width=120mm]{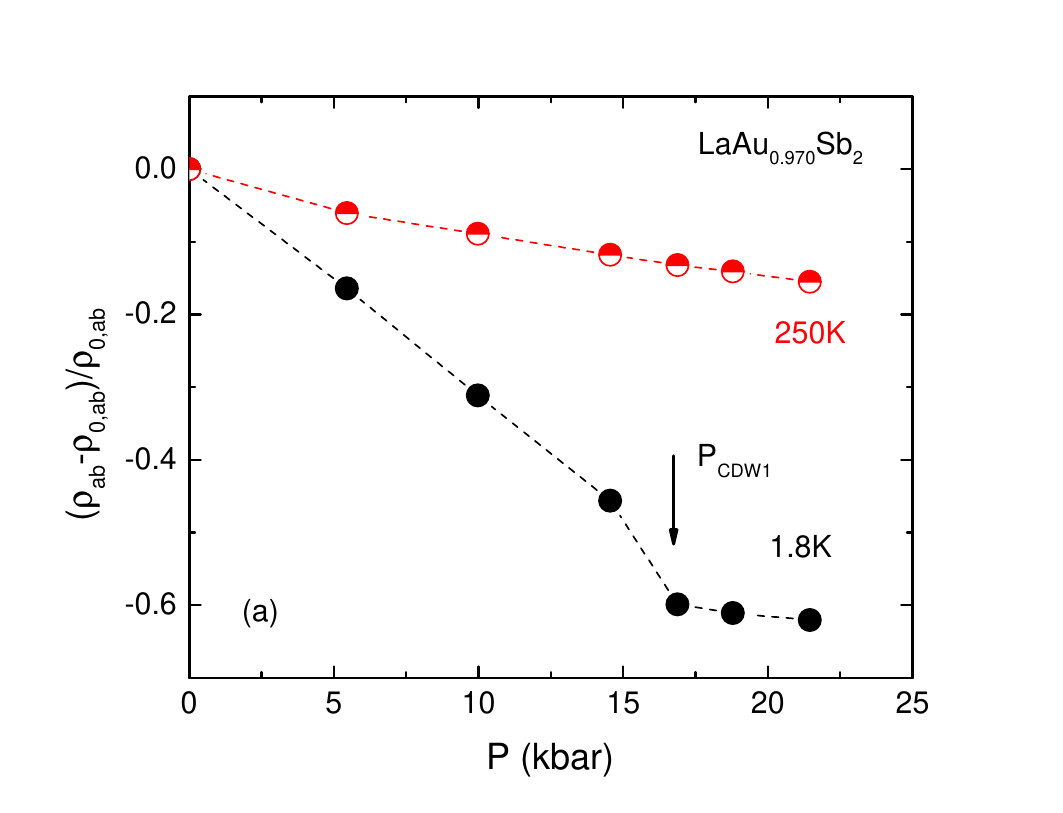}
\includegraphics[angle=0,width=120mm]{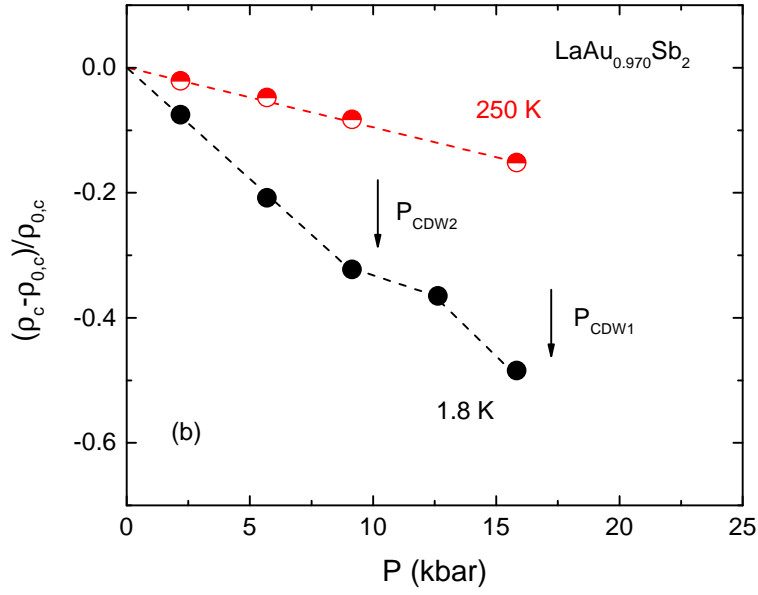}
\end{center}
\caption{(color online) Relative change of (a) in-plane and (b) $c$ - axis resistivity of  LaAu$_{0.970}$Sb$_2$ at 250~K and 1.8~K under pressure.  Arrow marks critical pressures for CDW1 and CDW2 suppression. } \label{F8}
\end{figure}

Pressure-induced relative changes of the in-plane resistivity of LaAu$_{0.970}$Sb$_2$ at the base temperature, 1.8~K, and above the CDW transitions, at 250~K, are presented in  Fig. \ref{F8}(a). At 250 K the resistivity decreases in an almost linear fashion, with a rate of $1/\rho_0~d\rho/dP = - 0.0070(4)$~1/K  which is close to $- 0.0088$~1/K reported for LaAgSb$_2$ at 300 K. \cite{tor07a} In contrast, the base temperature resistivity initially decreases 4-5 times faster, and then has a clear change of slope close to the critical pressure of CDW1 QCP. This is not unexpected, since below the critical pressure an additional contribution from suppression of the resistive increase associated with the CDW and its associated Fermi surface nesting plays an important role.

A similar set of data for the $c$ - axis resistivity is shown in Fig. \ref{F8}(b). The 250~K data show a close to linear decrease with a rate of $- 0.0097(4)$~1/K, that is not far from that for in-plane resistivity. Initially, the 1.8~K resistivity decreases  3-4 times faster, with a rate of $- 0.036(1)$~1/K. The 1.8~K data set has a clear anomaly close to $P_{CDW2}$, the critical pressure for CDW2. Unfortunately, the maximum pressure for the $c$ - axis resistivity run was below the  $P_{CDW1}$, so we were not able to evaluate if there is any anomaly associated with it.

The transitions from normal to CDW1 state as well as from CDW1 to CDW2 state appear to be of the second order, so the suppression of the CDW1 to $T = 0$~K could be recognized as a CDW quantum critical point. We further examine (magneto-) transport properties in the vicinity of the CDW QCP in some detail.

\begin{figure}
\begin{center}
\includegraphics[angle=0,width=140mm]{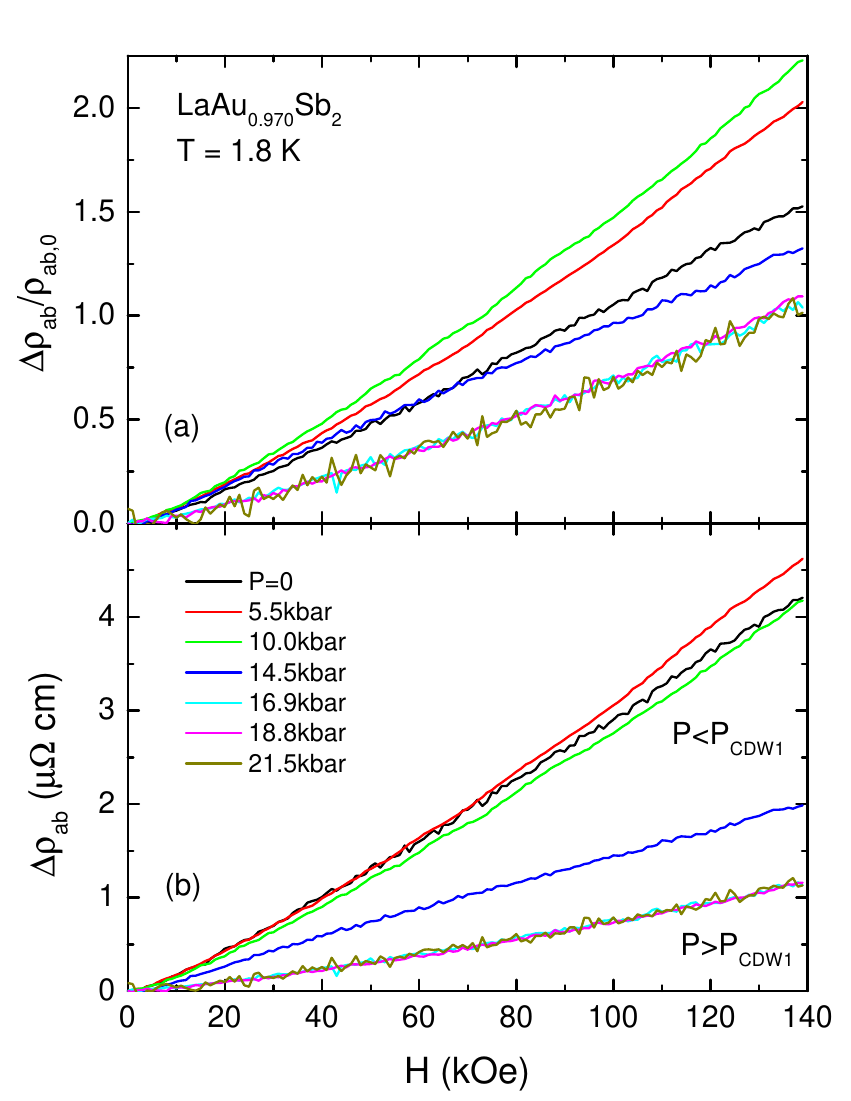}
\end{center}
\caption{(color online) (a) Transverse magnetoresistance, $\Delta \rho_{ab}/\rho_{ab,0} = [\rho_{ab}(H) - \rho_{ab}(H=0)]/\rho_{ab}(H=0)$, ($I || ab$, $H || c$), and  (b) change in resistivity in applied magnetic field, $\Delta \rho_{ab} = \rho_{ab}(H) - \rho_{ab}(H=0)$, ($I || ab$, $H || c$)   of  LaAu$_{0.970}$Sb$_2$ at 1.8~K measured under pressure up to 21.5~kbar.} \label{F9}
\end{figure}

Transverse magnetoresistance, $\Delta \rho_{ab}/\rho_{ab,0} = [\rho_{ab}(H) - \rho_{ab}(H=0)]/\rho_{ab}(H=0)$,  ($I || ab$, $H || c$)  of  LaAu$_{0.970}$Sb$_2$ was measured up to 140~kOe at 1.8~K (Fig. \ref{F9}(a)). It is non-saturating and at the maximum field has respectable values between $\sim 225$\% and  $\sim 100$ \%. The data for $P \geq 16.9$~kbar basically fall on the same line. If one re-plots these data as change of resistivity in applied field (without normalizing by the zero field resistivity, $\Delta \rho_{ab} = \rho_{ab}(H) - \rho_{ab}(H=0)$, Fig. \ref{F9}(b)) the results are even more curious: the data are separated into two well-defined manifolds: $P < P_{crit}$ ($P \leq 10.0$~kbar) and $P > P_{crit}$ ($P \geq 16.9$~kbar) with the data taken at the pressure close to critical, $P = 14.5$~kbar, being  in between these two manifolds.

\begin{figure}
\begin{center}
\includegraphics[angle=0,width=140mm]{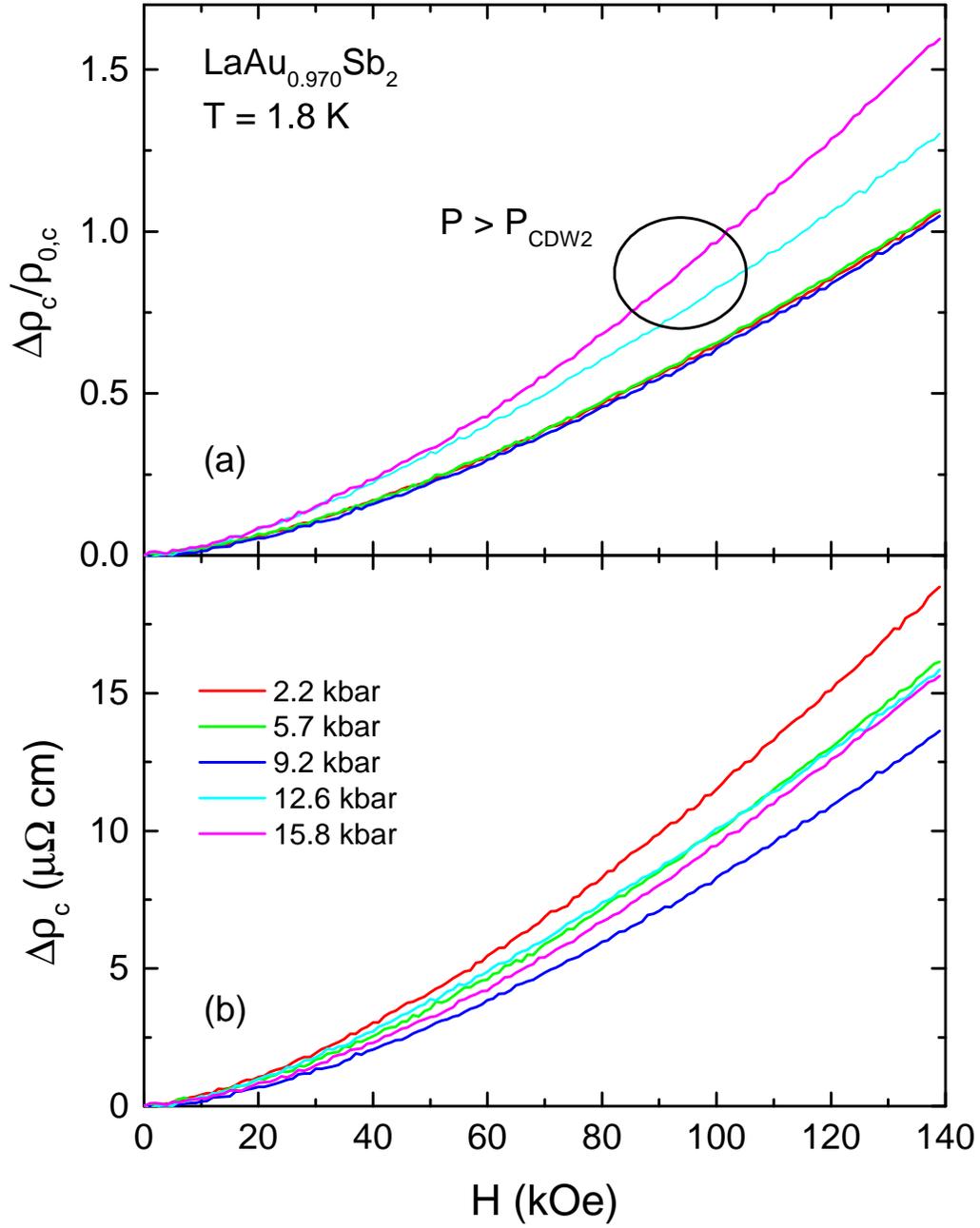}
\end{center}
\caption{(color online) (a) Transverse magnetoresistance ($I || c$, $H || ab$), and  (b) change in resistivity in applied magnetic field ($I || c$, $H || ab$)   of  LaAu$_{0.970}$Sb$_2$ at 1.8 K measured under pressure up to 15.8~kbar.} \label{F9.5}
\end{figure}

Similar data for $I || c$, $H || a$ are presented in Fig. \ref{F9.5}. Although there is not a clear segregation of the change in resistivity data, we can see that whereas the  field dependent magnetoresistance for $ P \leq P_{CDW2}$ scale to the same curve, the data for $ P >  P_{CDW2}$ appear to be clearly separate.

\begin{figure}
\begin{center}
\includegraphics[angle=0,width=140mm]{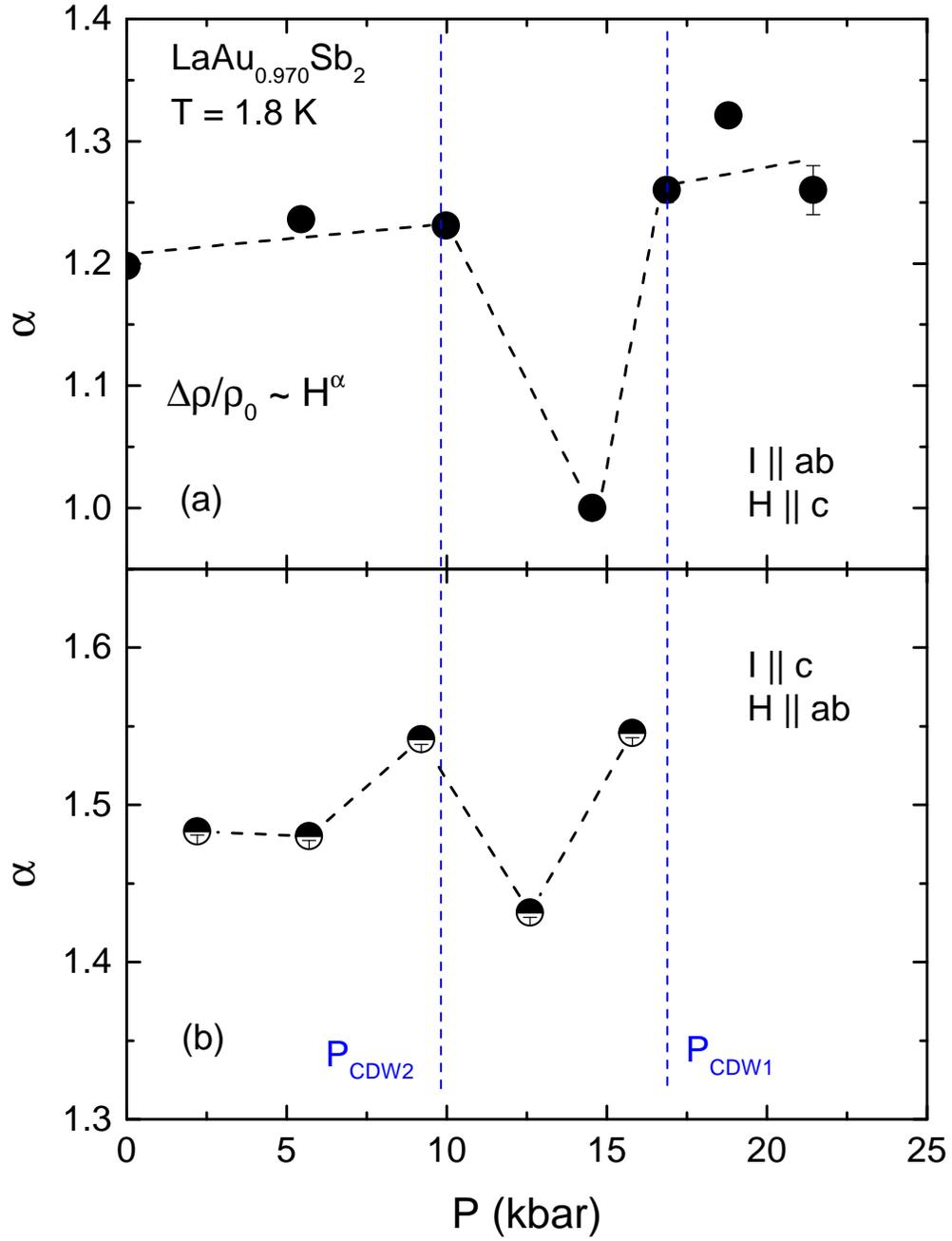}
\end{center}
\caption{Pressure dependence of the exponent $\alpha$ in magnetoresistance ($\Delta \rho/\rho_0 \propto H^{\alpha}$) (a) for $I || ab$, $H || c$, and (b) for  $I || c$, $H || ab$. Dashed lines are guide for the eye. Vertical dashed lines mark critical pressures for CDW1 and CDW2.} \label{F11}
\end{figure}

To quantify the evolution of the curvature of the field-dependent magnetoresistance with pressure we can re-plot the data from Fig. \ref{F9}(a) and Fig. \ref{F9.5}(a) on a {\it log - log} scale (see Fig. \ref{AMR} in the Appendix \ref{AB}) and perform linear fits of the data (between $\sim 20$~kOe and 140~kOe). Resulting slopes that are exponents $\alpha$ in $\Delta \rho/\rho_0 \propto H^{\alpha}$ are plotted as a function of pressure in Fig. \ref{F11} for in-plane and $c$ - axis resistivity data. Indeed, there is a clear change in the exponents $\alpha$ between the critical pressures for CDW2 and CDW1.

\begin{figure}
\begin{center}
\includegraphics[angle=0,width=140mm]{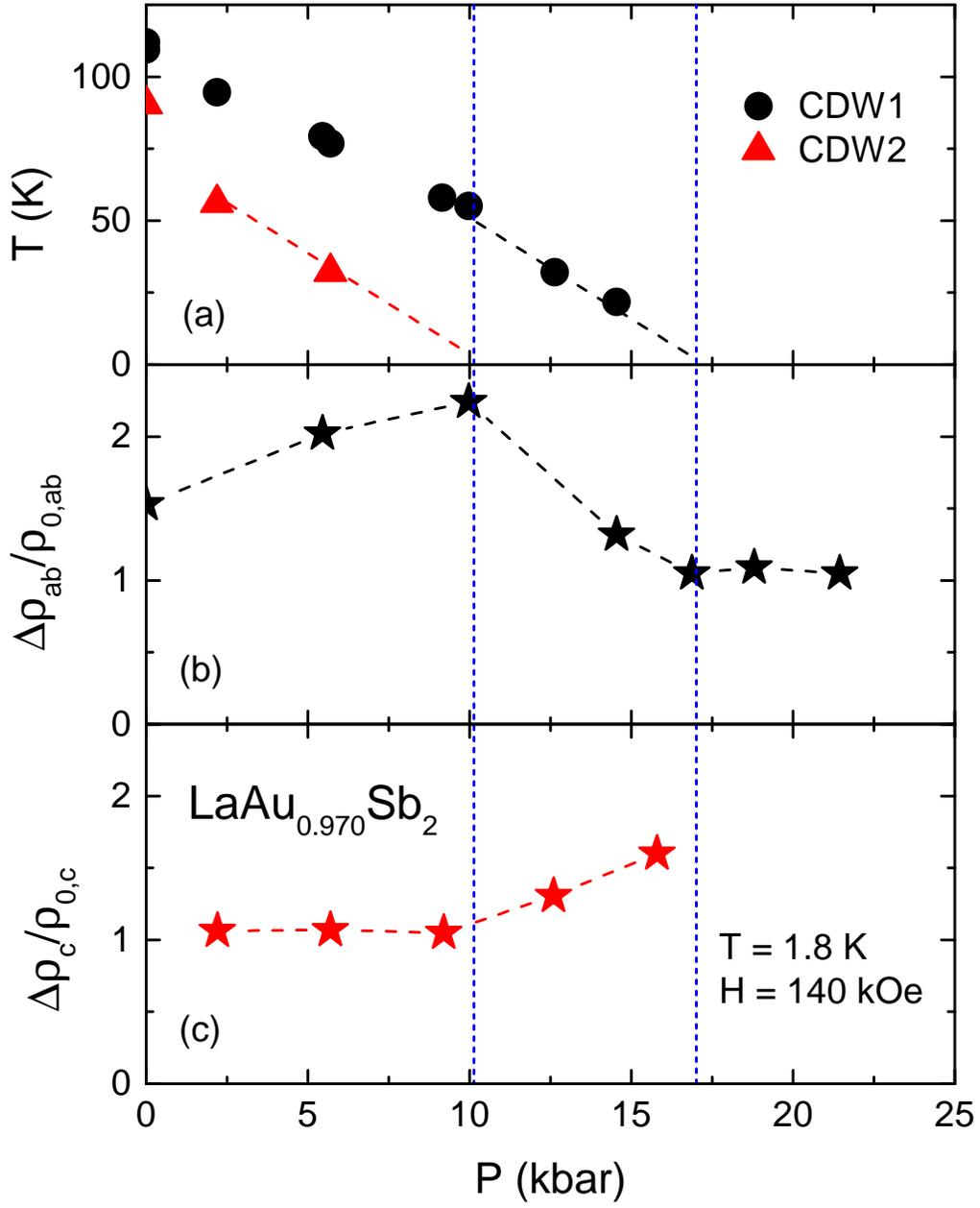}
\end{center}
\caption{(color online) Panel (a): $P~ - ~T$ phase diagram for  LaAu$_{0.970}$Sb$_2$. Panels (b) and (c): magnetoresistance at $T = 1.8$~K and $H = 140$~kOe as a function of pressure for $I || ab$, $H || c$ and  $I || c$, $H || ab$ respectively.   Dashed lines are guide for the eye. Vertical dashed lines mark critical pressures for CDW1 and CDW2.} \label{F12}
\end{figure}

Another parameter to follow is the value of magnetoresistance at 140~kOe as a function of pressure. This parameter entangles zero-field resistivity and the functional dependence of magnetoresistance on the applied field. Surprisingly, it appears that for $\rho_{ab}$ data this parameter displays anomalies associated with suppression of both CDW1 and CDW2, not just the dominant CDW1 (Fig. \ref{F12}(b)). For the $\rho_c$ data an anomaly at $T_{CDW2}$ is clearly seen (Fig. \ref{F12}(c)).

\section{Summary}

In this work we were able to tune and control Au occupancy in LaAu$_x$Sb$_2$ single crystals by changing the initial concentration of the elements in the melt.  The value of $x$ varied from from $x = 0.913(5)$ to $x = 0.991(7)$, using values of $x$ from Au site occupancy obtained in the Rietveld refinement of the powder x-ray data. For all the samples in this Au concentration range two CDWs were observed in the combination of in-plane and $c$-axis electrical transport. The CDW temperatures decrease monotonically from $T_{CDW1} = 110$~K and $T_{CDW2}  = 90$~K for $x = 0.991$ to  $T_{CDW1} = 33$~K and $T_{CDW2}  = 11.5$~K for $x = 0.913$. This behavior is in general agreement with the expected effect of non-magnetic impurities (or increase of non-magnetic scattering) on CDW discussed in literature, \cite{sch74a,bul74a,bul76a,gom84a} although a (small) change in band filling related to Au site occupancy could contribute to the change in CDW temperatures as well. 

The CDW temperatures are suppressed under pressure. For  LaAu$_{0.970}$Sb$_2$ a CDW QCP associated with the suppression of $T_{CDW1}$ to zero occurs at $\sim 17$~kbar. Anomalies in pressure dependence of the base temperature resistivity and transverse magnetoresistance (including via the exponent $\alpha$ of $\Delta \rho/\rho_0 \propto H^{\alpha}$) are observed at the CDW QCP. The in-plane magnetoresistance measured at 1.8~K and 140~kOe has clear anomalies at two critical pressure values, when either CDW1 or CDW2 are suppressed to $T = 0$~K. For the $c$ - axis magnetoresistance an anomaly at $P_{CDW2}$ is clearly observed, whereas $P_{CDW1}$ is beyond the pressure range of the measurements. The behavior of magnetoresistance at CDW QCP requires further experimental and theoretical studies. 

All in all, this work presents two ways of tuning charge density waves in an intermetallic crystals, opening the door for further, detailed studies of the CDW phenomenon.

\begin{acknowledgments}

The authors thank Elena Gati and Raquel A. Ribeiro for useful discussions, and Alexandra Elbakyan for assistance in literature searches. Work at the Ames Laboratory was supported by the U.S. Department of Energy, Office of Science, Basic Energy Sciences, Materials Sciences and Engineering Division. The Ames Laboratory is operated for the U.S. Department of Energy by Iowa State University under contract No. DE-AC02-07CH11358. LX was supported, in part, by the W. M. Keck Foundation. Much of this work was carried out while DHR was on sabbatical at Iowa State University and Ames Laboratory and their generous support (again under under contract No. DE-AC02-07CH11358) during this visit is gratefully acknowledged. DHR was supported as well by Fonds Qu\'eb\'ecois de la Recherche sur la Nature et les Technologies.

\end{acknowledgments}

\appendix

\section{Rietveld Refinement and EDS Results}
\label{AA}
\setcounter{table}{0} \renewcommand{\thetable}{A\arabic{table}} 

This appendix contains tables with the results of Rietveld refinements and EDS chemical analysis of five LaAu$_x$Sb$_2$ samples.

\begin{table*}[h]
\begin{ruledtabular}
\caption{\label{T1}Lattice parameters of LaAu$_x$Sb$_2$ samples grown using different starting compositions\\}
\begin{tabular}{cccc}
Sample&$a$ (\AA)&$c$ (\AA)&$V$ (\AA$^3$)\\ \hline
Au1&4.4475(1)&10.3476(6)&204.68(1)\\
Au2&4.4430(2)&10.4237(4)&205.77(1)\\
Au4&4.4358(1)&10.4552(3)&205.72(1)\\
Au6&4.4347(1)&10.4653(3)&205.81(1)\\
Au8&4.4341(1)&10.4718(4)&205.88(1)\\
\end{tabular}
\end{ruledtabular}
\end{table*}

\begin{table*}[h]
\caption{\label{T2}Atomic coordinates, occupancy, and isotropic displacement parameters of LaAu$_x$Sb$_2$ samples grown using different starting compositions\\}
\begin{ruledtabular}
\begin{tabular}{cccccccc}
Sample&atom&site&x&y&z&occupancy&$U_{eq}$\\ \hline

Au1&La&2c&0.25&0.25&0.2496(2)&1&0.0340(7)\\
&Au&2b&0.75&0.25&0.5&0.913(5)&0.0378(6)\\
&Sb1&2a&0.75&0.25&0&1&0.0378(6)\\
&Sb2&2c&0.25&0.25&0.6664(2)&1&0.0378(6)\\ \hline

Au2&La&2c&0.25&0.25&0.2488(3)&1&0.0288(7)\\
&Au&2b&0.75&0.25&0.5&0.936(6)&0.0311(8)\\
&Sb1&2a&0.75&0.25&0&1&0.0309(7)\\
&Sb2&2c&0.25&0.25&0.6693(2)&1&0.0309(7)\\ \hline

Au4&La&2c&0.25&0.25&0.2478(2)&1&0.0334(7)\\
&Au&2b&0.75&0.25&0.5&0.947(6)&0.0331(6)\\
&Sb1&2a&0.75&0.25&0&1&0.0344(6)\\
&Sb2&2c&0.25&0.25&0.6700(2)&1&0.0344(6)\\ \hline

Au6&La&2c&0.25&0.25&0.2465(2)&1&0.0337(6)\\
&Au&2b&0.75&0.25&0.5&0.970(5)&0.0346(5)\\
&Sb1&2a&0.75&0.25&0&1&0.0355(5)\\
&Sb2&2c&0.25&0.25&0.6703(2)&1&0.0355(5)\\ \hline

Au8&La&2c&0.25&0.25&0.2475(2)&1&0.0252(7)\\
&Au&2b&0.75&0.25&0.5&0.991(7)&0.0282(7)\\
&Sb1&2a&0.75&0.25&0&1&0.0252(6)\\
&Sb2&2c&0.25&0.25&0.6704(2)&1&0.0252(6)\\
\end{tabular}
\end{ruledtabular}
\end{table*}

\begin{table*}[h]
\begin{ruledtabular}
\caption{\label{T3}EDS results for LaAu$_x$Sb$_2$ samples grown using different starting compositions\\}
\begin{tabular}{ccccccc}
Sample&La at.\%&Au at.\%&Sb at. \%&Au/La&2Au/Sb&3Au/(La+Sb)\\ \hline
Au1&26.41(9)&23.09(4)&50.64(8)&0.874(4)&0.912(3)&0.899(4)\\
Au2&26.1(2)&23.9(1)&50.1(1)&0.92(1)&0.954(7)&0.941(9)\\
Au4&25.8(1)&24.55(4)&49.61(7)&0.951(5)&0.982(4)&0.971(5)\\
Au6&25.9(2)&24.55(4)&49.61(7)&0.949(8)&0.990(3)&0.976(6)\\
Au8&25.71(4)&24.65(7)&49.70(7)&0.959(4)&0.992(4)&0.981(7)\\
\end{tabular}
\end{ruledtabular}
\end{table*}

\clearpage

\section{Magnetoresistance of L\lowercase{a}A\lowercase{u}$_{0.970}$S\lowercase{b}$_2$ under pressure}
\label{AB}

Fig. \ref{AMR} presents  transverse magnetoresistance $[\rho(H) - \rho(H = 0)]/\rho(H = 0$   of  LaAu$_{0.970}$Sb$_2$ at 1.8~K measured (a) for $I || ab$, $H || c$ under pressure up to 21.5~kbar, and (b) for $I || c$, $H || ab$ under pressure up to 15.8~kbar  plotted on a {\it log - log} scale. The slopes obtained in linear fits of these data (dashed lines) give the values of the exponent $\alpha$ in $\Delta \rho/\rho_0 \propto H^{\alpha}$. The fits were performed in the magnetic field range between $\sim 20$~kOe and 140~ kOe.

\section{L\lowercase{a}A\lowercase{u}$_{0.936}$S\lowercase{b}$_2$ under pressure}
\label{AC}
\begin{figure}
\begin{center}
\includegraphics[angle=0,width=140mm]{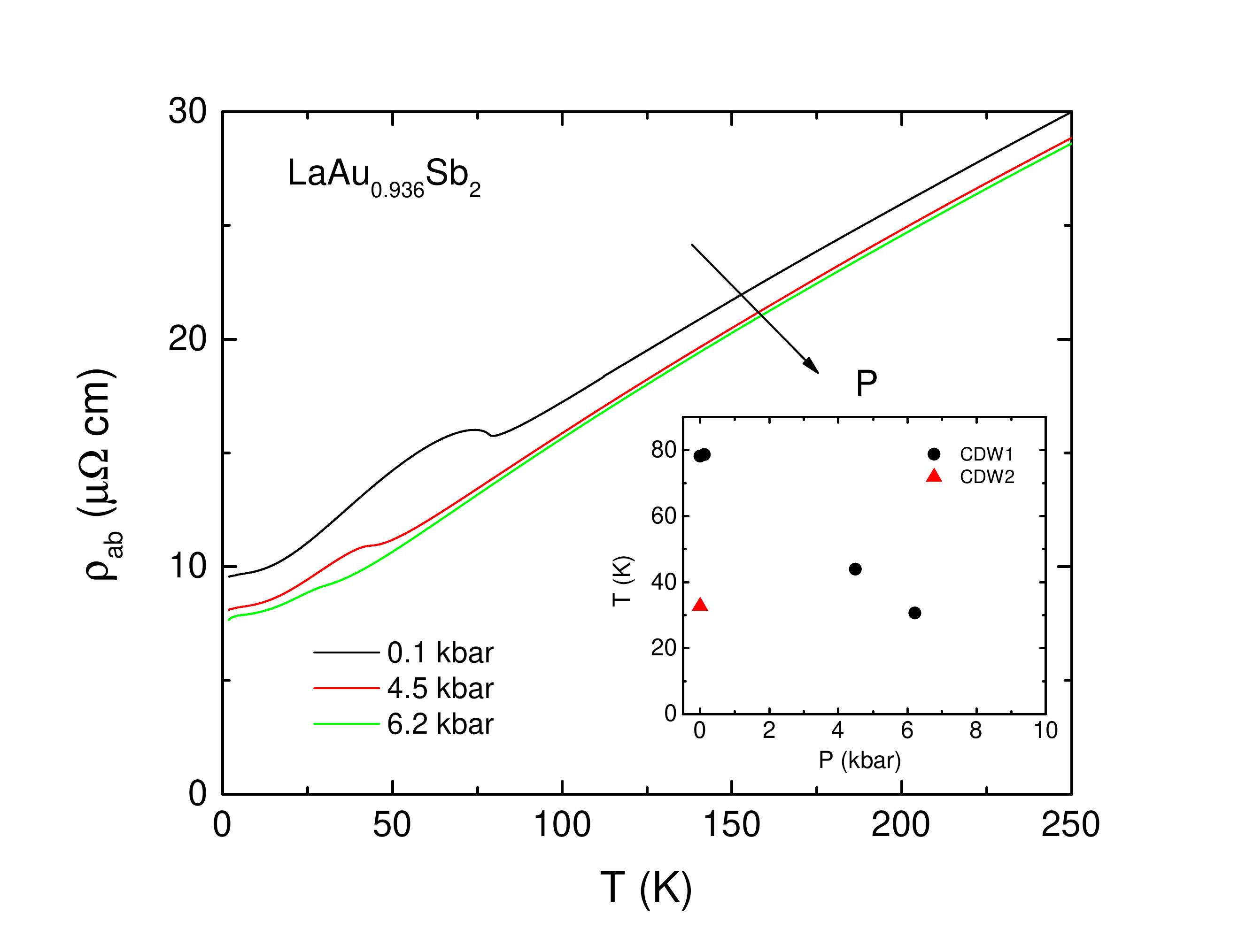}
\end{center}
\caption{(color online)  In-plane resistivity  of  LaAu$_{0.936}$Sb$_2$  under pressure. Arrow points to the direction of pressure increase. Inset: change of CDW temperatures under pressure.} \label{ARP}
\end{figure}

In plane resistivity under pressure has been measured for  LaAu$_{0.936}$Sb$_2$ for comparison to earlier literature as well as for comparison to the LaAu$_{0.970}$Sb$_2$ shown in the main text. The results are presented in Fig. \ref{ARP}. As for  LaAu$_{0.970}$Sb$_2$, resistivity decreases under pressure  and the CDW1 transition is suppressed. We were not able to follow the CDW2 under pressure most probably because it was already suppressed to $T = 0$~K by 4.5~kbar. The pressure derivative of $T_{CDW1}$ is $- 7.8(1)$~K/kbar, so the suppression of the $T_{CDW1}$ is faster than in the case of  LaAu$_{0.970}$Sb$_2$ and  LaAgSb$_2$. Linear extrapolation suggests that the critical pressure for CDW1 in  LaAu$_{0.936}$Sb$_2$ is $\approx 10$~kbar.

\begin{figure}
\begin{center}
\includegraphics[angle=0,width=140mm]{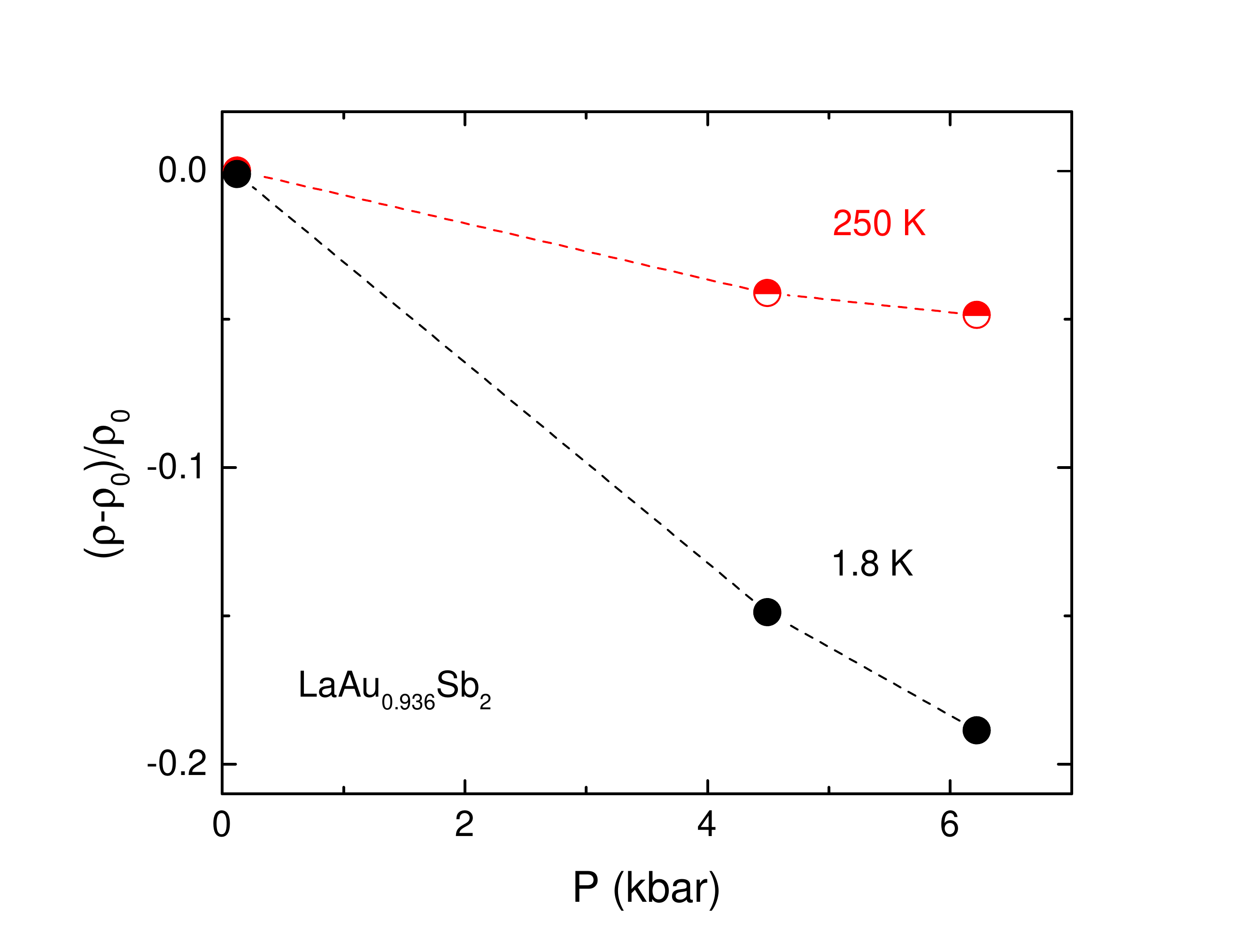}
\end{center}
\caption{(color online)  Relative change of in-plane resistivity of  LaAu$_{0.936}$Sb$_2$ at 250~K and 1.8~K under pressure. } \label{ARP1}
\end{figure}

Effect of pressure on the resistivity of  LaAu$_{0.936}$Sb$_2$ at the base temperature, 1.8~K, and above the CDW transitions, at 250~K, is shown in  Fig. \ref{ARP1}. At 250~K resistivity decreases with the rate of $1/\rho_0~d\rho/dP = - 0.008(1)$~1/K  which is close to the data for LaAgSb$_2$  \cite{tor07a} and LaAu$_{0.970}$Sb$_2$ in the same temperature range. Similar to  LaAu$_{0.970}$Sb$_2$, resistivity measured at 1.8~K initially is decreasing significantly faster ($1/\rho_0~d\rho/dP = - 0.031(2)$~1/K) than at 250~K.

\clearpage

\clearpage

\clearpage

\clearpage

\clearpage

\clearpage

\clearpage

\clearpage

\clearpage

\clearpage

\clearpage

\clearpage

\clearpage

\begin{figure}
\begin{center}
\includegraphics[angle=0,width=120mm]{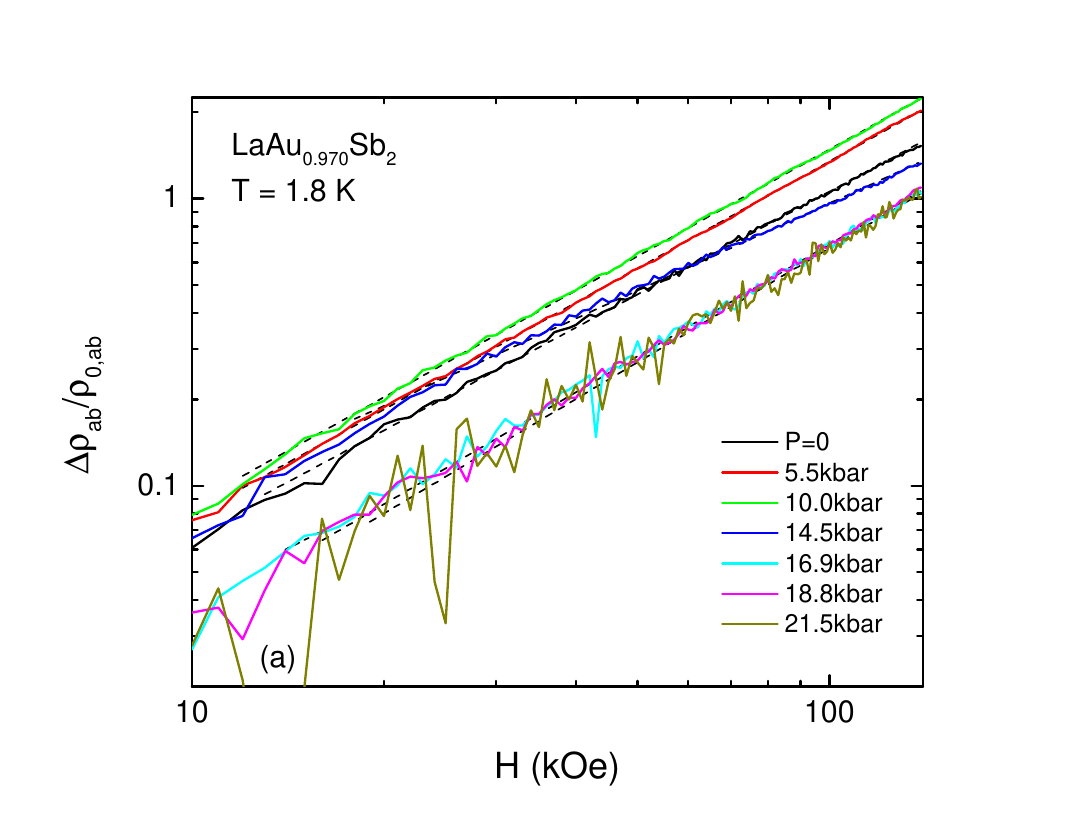}
\includegraphics[angle=0,width=120mm]{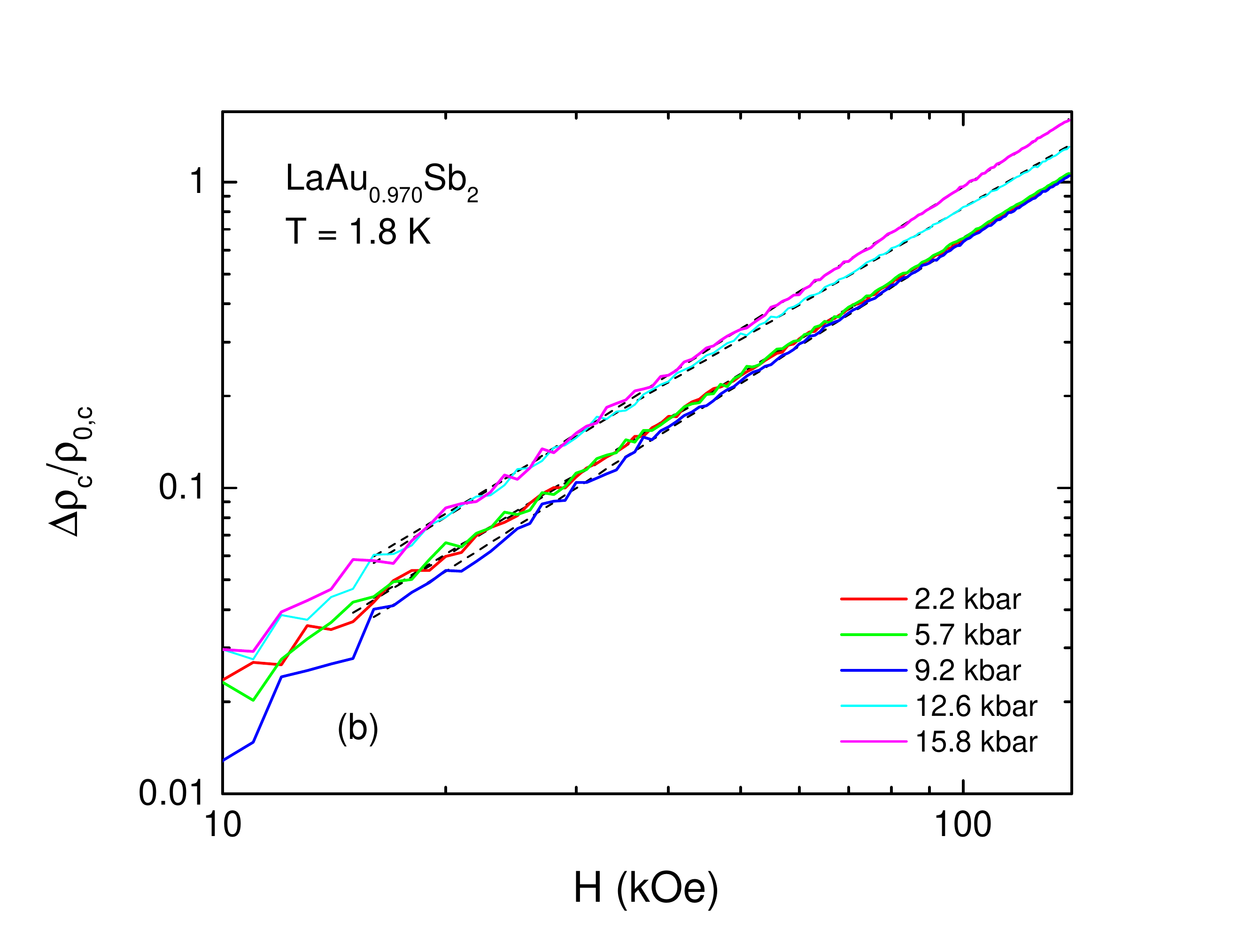}
\end{center}
\caption{(color online) (a) Transverse magnetoresistance  of  LaAu$_{0.970}$Sb$_2$ at 1.8 K plotted on a {\it log - log} scale (a) $I || ab$, $H || c$  measured  up to 21.5~kbar, (b) $I ||c$, $H || ab$   up to 15.8~kbar. Dashed lines are linear fits.} \label{AMR}
\end{figure}

\clearpage

\clearpage

\end{document}